\documentclass{article}
\usepackage[T1]{fontenc}
\usepackage[latin1]{inputenc}
\usepackage{float}
\usepackage[dvips]{graphicx}
\usepackage{amssymb}
\usepackage{setspace}
\begin{document}

\title{Monte Carlo study of the phase transition in the
Critical behavior of  the Ising model with shear }

\author{G. P. Saracco$^{1}$ and G. Gonnella$^{2}$}
\date{}
\maketitle
\noindent \emph{(1) Instituto de Investigaciones Fisicoqu\'{i}micas  y Aplicadas
(INIFTA), UNLP, CCT La Plata - CONICET, Casilla de Correo 16 Sucursal 4
(1900) La Plata, Argentina.\\
(2) Dipartimento di Fisica, Universit\`a di Bari and Istituto
Nazionale di Fisica Nucleare, Sezione di  Bari, via Amendola 173,
70126 Bari, Italy.}

\begin{abstract}

The critical behavior of the Ising model with non-conserved dynamics
and an external driving field
 mimicking a shear profile  is analyzed by studying  its dynamical evolution in the short time regime.
 Starting from high temperature disordered configurations (FDC),
 the critical temperature $T_c$ is determined when the order parameter,
 defined as the absolute value of the transversal spin profile,
 exhibits a power-law behavior with an exponent that is a combination of some of the
 critical exponents of the transition.
 For each value of the shear field magnitude, labeled as $\dot{\gamma}$,
  $T_c$ has been  estimated and two stages have been  found: 1) a growing stage at low values
  of $\dot{\gamma}$, where $T_c\sim\dot{\gamma}^\psi$ and   $\psi=0.52(3)$;
 2) a saturation regime at large $\dot{\gamma}$.
 The same values of $T_c(\dot{\gamma})$ were found studying the dynamical evolution
   from the ground state configuration  (GSC)
  with all spins pointing in the same direction.
   By combining the exponents of the corresponding power laws obtained from each initial configuration
   the set of critical exponents was calculated.
   These values, at large external field magnitude, define a new critical behavior
   different from that  of the Ising model
   and  of other  driven lattice gases.
   %Furthermore, we also show that the longitudinal correlations, i.e along the field axis,
   %are more relevant than the transversal ones in the critical dynamic behavior of the model.
   \\
\end{abstract}
%\pacs{PACS numbers:  64.75.+g; 05.70.Ln; 47.20.Hw}

\section{Introduction}
\label{1} The statistical mechanics of equilibrium phenomena is a very useful theoretical
framework for understanding  the thermodynamic properties of many--particle systems from a microscopical
point of view. However, in nature, most of the systems evolve under
out-of-equilibrium conditions, and  there is not yet
a suitable general framework to study them as in the case of equilibrium systems.
Nevertheless, some progress have been achieved in the knowledge of  far from equilibrium behavior by means
 of simple models, capable to catch the essential physics of non-equilibrium processes.\\
 In this context, we introduce a very simple model, derived from the Ising model,
  driven out of equilibrium by an external field that  mimics the effects of a uniform shear profile \cite{B94}.
  This  model evolves with a non-conserved dynamics, corresponding to model A in the classification of Hohenberg
   and Halperin \cite{HH}, and it was already used by Cavagna \textsl{et. al} \cite{CBT00} and by
   Cirillo \textsl{et. al} \cite{ciri} to study  phase separation. \\

We will focus on the study of phase transition properties in this
model. Typical configurations observed in our simulations are
displayed in figure {\ref{snap}. At low temperatures the system
appears ordered with elongated domains directed along the field
direction. At high temperatures the system exhibits a gas-like
appearance with disordered patterns. Similar ordered and disordered
phases, also experimentally found \cite{larson}, generally
characterize the behavior of sheared binary systems. As  usually in
systems with an applied external field, the transition point is a
function of the magnitude of the driving field \cite{onuki}.
\begin{figure}[H]
\centering
\includegraphics[height=7cm,width=11cm,clip=,angle=0]{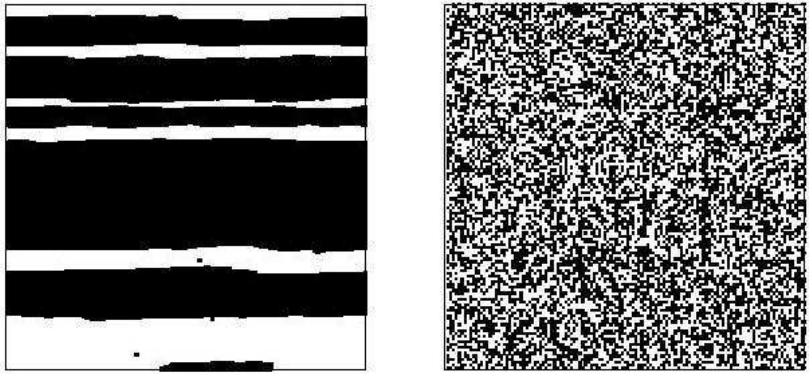}
\caption{Snapshot configurations corresponding to the two  phases of
the Ising model with shear. On the left  panel we observe a
 typical stripe-like configuration at $T<T_c$. On the right  panel a configuration at $T>T_c$ is displayed.
  The external field is applied along the horizontal axis.}\label{snap}
\end{figure}

Previous theoretical and experimental  studies have  shown  that
sheared binary systems undergo a second order phase transition at a
critical temperature $T_c$ \cite{onuki}. In diffusive systems the
effect of the external driving field is to inhibit fluctuations so
that the critical temperature is expected to increase with the
magnitude of the driving. In a continuum model with non--conserved
dynamics, in the large-$N$ analytical approximation, it has been
found that the value of the critical temperature depends on the
driving field following a power law at small field magnitudes
\cite{gonepeli}. In previous Monte Carlo studies on the  critical
behavior of sheared Ising models \cite{chan}, it was not possible to
extract information about the critical temperature, due to numerical
uncertainties and finite size effects \cite{prl2}. In view of this,
we revisit this issue, in order to determine the critical
temperature of the model as a function of the magnitude of the
external field, and to compute for the first time the critical
exponents in this model. For the sake of comparison, we will
contrast the obtained values  with those computed for the 2d driven
lattice gas model (DLG) \cite{kls} that will be briefly described in
the next section.

To study the phase transition in the model, the critical dynamical
behavior will be investigated by monitoring the time evolution of some observables before
 the system reaches nonequilibrium steady states (NESS). This technique, generally called
  \textsl{short time dynamics} \cite{JSS, zhe2} is an alternative  convenient way to
 obtain both the critical temperature and exponents precisely, with less computational
 cost than other methods commonly used, such as the finite size scaling applied to the
 specific heat and response functions, that require a considerably amount of simulation time
  in order to reach NESS. Furthermore, since the measurements are carried out
   in the first steps of evolution, the short time dynamic approach is free of  the critical slowing down.\\

The manuscript is organized as follows: In Section \ref{2}, the Ising model with the external shear field is introduced. The technique used to study the phase transition is described in Section \ref{3}. The simulation results are presented in Section \ref{4}, and finally the conclusions are stated in Section \ref{5}.

\section{The model}
\label{2}
We will consider the nearest--neighbor two--dimensional
Ising model with a single--spin--flip thermalization dynamics,
e.g.\ the Metropolis dynamics  \cite{metro}. The driving field will be defined
in order to mimic the convective velocity shear profile
\begin{equation}
\label{sh} v_x(y)=\dot{\gamma} y \qquad v_y=0
\end{equation}
where the  parameter $\dot{\gamma}$ is called the {\em shear
rate} and represents the shear field magnitude. If the system is imagined as a sequence of layers
labelled by $y$, then $\dot{\gamma} y$ is the displacement of the
layer $y$ in a unit of time. If $L_y$ is the vertical size and
$v_{\textrm{max}}$ is the speed of the fastest layer, then
$\dot{\gamma} L_y=v_{\textrm{max}}$.

The model is defined on a square lattice $\Lambda$ of horizontal and vertical sizes $L_x$, $L_y$ respectively, with periodic boundary conditions in the $L_x$ direction and free in the $L_y$ direction. More precisely, let $\Omega=\{-1,+1\}^\Lambda$ be the space of configurations and, for $\sigma\in\Omega$, let $\sigma_{x,y}$ be the value of the spin associated to the site
$(x,y)\in\Lambda$. Then the Hamiltonian of the model is
\begin{equation}
\label{ham} H_\Lambda(\sigma)=
-J\sum_{y=1}^{L_{y}}\sum_{x=1}^{L_{x}}\sigma_{x,y}\sigma_{x+1,y}-
J\sum_{x=1}^{L_{x}}\sum_{y=1}^{L_{y}-1}\sigma_{x,y}\sigma_{x,y+1}
\end{equation}
with $\sigma_{L_{x}+1,y}=\sigma_{1,y}$ for all $y=1,\dots,L_{y}$,
and $J$ is a positive real coupling constant, which means that the
interactions are ferromagnetic. We will  combine the thermalization
dynamics with an algorithm introducing the shear in the system. The
shear is superimposed to the thermalization dynamics with typical
rates not depending on the thermalization phenomenon, but fixed a
priori. This has been done in different ways in \cite{CBT00, chan,
okabe}.  In this paper we use a very ductile generalization of those
dynamics aiming to introduce the shear effects in a way resulting
competitive with respect to the thermalization process.
Notice that our dynamics results from the  combination of two steps:
i) a thermalization  step which would bring the system in the
usual  equilibrium; ii)  a shear step which changes the
configurations of the system  forbidding to reach the equilibrium.
 Therefore, all together, our algorithm  does not satisfy
 local detailed balance expressed in terms of standard
  equilibrium probabilities of configurations. Similar models
  have been also used in different context of non-equilibrium studies \cite{crooks}.\\

 Let the {\em
time unit} be the time needed for a full thermal update of the
entire lattice, e.g.\ a full sweep of the Metropolis algorithm. The
shear algorithm is parametrized with a submultiple $\tau$ of
$L_{y}L_{x}$ (the period of the shear procedure), a positive integer
$\lambda\le L_{x}/2$ (the number of unit cells that a row is shifted when the shear is
performed), and a non--negative real $\nu \le1/L_{y}$. The dynamics
of the model that we study in this paper is defined in a precise way
via the following algorithm:
\begin{enumerate}
\item \label{i:tre.1}
  set $t=0$, choose $\sigma_0\in\Omega$, and set $n=0$;
\item \label{i:tre.2}
 increase by 1 the index $n$, and
  choose at random with uniform probability $1/L_{x}L_{y}$ a site
  of the lattice and
  perform the elementary single--site step of the thermalization dynamics;
\item \label{i:tre.3}
  if $n$ is multiple of $\tau$  a layer is
  randomly chosen with uniform probability $1/L_{y}$. Then, if $\bar y$ is the chosen layer,
  all the layers with $y\ge\bar y$ are shifted by $\lambda$ lattice
  spacings to the right with probability $\nu L_{y}$;
\item \label{i:tre.3.1}
 if $n<L_{x}L_{y}$ goto~\ref{i:tre.2}, else denote by $\sigma_{t+1}$ the configuration of the system;
\item \label{i:tre.4}
  set $t=t+1$, set $n=0$, and goto~\ref{i:tre.2}.
\end{enumerate}
\par\noindent
We note that if $\nu=1/L_{y}$ the shift at step \ref{i:tre.3} is surely performed and  this
case will be later addressed to as {\it full shear}. The smoothness of the shear field, eqs.
(\ref{sh}), is ensured by the random choice of the layer $\bar{y}$ in step 3.
Now we want to express the shear rate $\dot{\gamma}$, introduced in equation (\ref{sh}), in terms of the
parameters of our dynamics. We have to estimate the typical displacement per unit of time of
the row labelled by $y$. Such a row is involved in a shear event, step~\ref{i:tre.3} of the
algorithm above, if and only if the extracted row $\bar y$ is such as $\bar y\le y$, and
this happens with probability $y/L_{y}$. Since the shear event results in a shift with
probability $\nu L_{y}$, the probability that during a shear event the row $y$ does shift is
given by
\begin{displaymath}
\frac{y}{L_{y}}\times\nu L_{y}=\nu y.
\end{displaymath}
By noting that the number of shift events per unit of time is
equal to $L_{x}L_{y}/\tau$ and recalling that the shift amplitude
is $\lambda$, we have that the typical shift of the row $y$ per
unit of time is given by
\begin{displaymath}
\frac{L_{x}L_{y}}{\tau}\times\lambda\times\nu y.
\end{displaymath}
By using definition (\ref{sh}) we finally get $ \dot{\gamma}=L_{x}L_{y}\nu \lambda/\tau$,
which becomes $\dot{\gamma}=L_{x}\lambda/\tau$ in the case of full shear. \\
It is important to remark that there exists a large variety of
models that evolve under non equilibrium states  by the action of an
external field. An example is the driven
 lattice gas (DLG) where  the driving field is not
 superimposed to the thermalization dynamics, but it is rather inserted in the Metropolis
 transition rates, that become anisotropic, and biases the movement of particles along
 its direction \cite{kls}.
Furthermore, it exhibits a second order phase transition at particle density $\rho=1/2$, between an \textsl{ordered} phase at low temperatures characterized by regions of low and high particle density,
called stripes, oriented along the field direction, and a \textsl{disordered} phase
at high temperatures with the appearance of a lattice gas. Both ordered and disordered phases have a similar appearance with those exhibited in figure \ref{snap}.
The critical temperature
increases with the magnitude of the external field, saturating at $T_c\sim1.41 T_c(0)$
in the case of infinite \cite{kls}.  Here, $T_c(0)=2.269$ $J/k_B$ is the critical temperature of the
2d Ising model ($J$ is the coupling constant and $k_B$ is the Boltzmann constant).

\section{Dynamical Critical Behavior}
\label{3} It is known that, for systems exhibiting critical behavior, the relevant observables
measured in equilibrium stationary states can be written in terms of power laws, with
characteristic critical exponents due to the divergence of  both the spatial correlations  and the correlation time.
 In recent years, however, the attention has been also focused to the \textsl{early stages of the
evolution} of the system towards the critical state, that is,
 to a microscopic time regime where the spatial correlation length is small compared with
the system size \cite{JSS}. Within this regime, it is possible to
measure scaling laws of the observable quantities \cite{zhe2, zhe}.
This new method to study second-order phase transitions, called
\textsl{short time dynamics}, allows to estimate the critical
temperature and to compute the critical exponents of the transition
with relative quickness and avoids the shortcomings that more usual
techniques present to study critical behavior. Furthermore, the
short time dynamics has been applied to investigate the critical
behavior of a wide range of systems of different nature, such as
models showing criticality under equilibrium conditions, such as
e.g, the XY systems \cite{trimper}, the 2d 3-state Potts model
\cite{zhe21}, the Ising magnet under different lattice geometries
\cite{zhe3, bab}, and of nonequilibrium critical models such as the
driven diffusive lattice gas (DLG) \cite{alsa, repalsa, saal}, etc.
 In these
last three works, a detailed analysis of the second-order \cite{alsa, repalsa}
and first-order \cite{saal} non-equilibrium phase transition was
performed by using the short time critical dynamic methodology.  For
the second-order phase transition, the excellent agreement between
critical exponents evaluated using the standard (stationary) and
dynamical (short time) approaches strongly support the robustness of
this method. Encouraged by this success, our goal is to extend the
short time dynamics concept to this model, basing our ideas
 on the already developed short time dynamics method for the DLG model in ref. \cite{alsa}.\\
The above mentioned scaling laws can be observed employing two different initial configurations,
namely: 1) Fully disordered configurations (FDC's), which means that the system is initially
placed in a thermal bath at $T \rightarrow \infty$ , and the system configuration is similar to that exhibited on the right panel of figure \ref{snap}; 2) Completely ordered configurations or
ground state configurations (GSC's) as expected for $T = 0$. In our model, based on the fact
that the equilibrium Ising model has all the spins pointing in the same direction (i. e.
magnetization $M=$ 1 or -1) at this temperature, we will adopt this configuration as the
ground state for testing the short time dynamic behavior. \\
The shear field introduces anisotropic effects, that generates
anisotropic correlations in the system. As a consequence of this,
there will be two correlation lengths, namely: 1) A
\textit{parallel} or longitudinal correlation length
$\xi_{\parallel}$
  along the  external field direction, and 2) A \textit{perpendicular} or transverse
   correlation length $\xi_{\perp}$  perpendicular to $\xi_{\parallel}$.  Whatever
    the initial condition is used to start the system, both spatial correlation lengths are
     quite small or zero at the beginning of the dynamic process, and near the critical temperature $T_c$ they increase dynamically as a power law $\xi_{\parallel(\perp)}\propto t^{1/z_{\parallel(\perp)}}$, where $z_{\parallel(\perp)}$ is the dynamic critical exponent in the respective directions. \\
Before we start to describe the theoretical basis of the technique
applied to this model, we set  the external driving field  along the
horizontal direction, i.e. the $x$ axis. Also, we need to define
quantities that are relevant in the critical behavior of the model.
Based on the  morphological appearance of typical configurations
present in the
 system (see figure \ref{snap}), we will consider a variant of the order parameter $OP$ employed
in the critical study of the DLG model \cite{alsa}:
\begin{equation}
\label{op}
OP=\frac{1}{L_y}\sum_{y=1}^{L_{y}} |P(y)|,
\end{equation}
\noindent where
$P(y)=\frac{1}{L_x}\sum_{x=1}^{L_{x}} \sigma_{x,y}$ is the average
of the spin profile in the shear field axis.

\noindent The order parameter defined in this way can take into account the
small ordering that appears at the early
stages of the evolution.\\
There is one more point to take into account before we start to
 expose the method applied to this system. In all formulas below we will assume, and demonstrate later,
 that only the parallel correlation length is relevant in the short time critical evolution of the system.
 In fact at $T\simeq T_c(\dot{\gamma})$ and at early times of evolution, parallel and perpendicular
 correlations begin to increase.  However,  domains of perpendicularly
  correlated spins are broken by the shear, and assume a  characteristic elongated  shape,
   also observed in many experimental studies of sheared systems.
 As a consequence of
  this, transversal correlations grow slower than parallel correlations, so they do not take part in
  the dynamic critical behavior of the model at short times.  This effect
    was also  shown for the DLG model \cite{alsa, repalsa}.  Since
this happens independently of the initial
  configuration, we will take $z=z_{\parallel}$ in every expression below.
Furthermore, they  must contain the anisotropic finite size dependence in order to match the usual anisotropic scaling forms for the NESS regime.\\
Starting with FDC's, the scaling law proposed for the order parameter ($OP$) reads \cite{alsa},
\begin{equation}
\label{2op}
OP(t,\phi,L_x,L_y)=b^{-\beta/\nu_{\parallel}}
 OP^{*}(b^{-z}t,b^{1/\nu_{\parallel}}\phi, b^{-1}L_{x},b^{-\nu_{\perp}/\nu_{\parallel}}L_y) ,
\end{equation}

\noindent where $t$ is the time, $b$ is the spatial rescaling factor, $\beta$ is the
critical exponent of the order parameter, $\nu_{\parallel(\perp)}$
are the correlation length critical exponents in the $x$($y$)
axis ($\xi_{\parallel(\perp)}\propto\phi^{-\nu_{\parallel(\perp)}}$), $OP^{*}$ is a scaling
function, $z$ is the already mentioned dynamic exponent of the longitudinal correlation length,
 and $\phi=\frac{T-T_{c}}{T_{c}}$ . Notice also that $L_y$ is
 rescaled by $b^{-\nu_{\perp}/\nu_{\parallel}}$ to include possible shape effects
 in the dynamic critical behavior \cite{kls}.\\
To generate the FDC initial conditions, the lattice is filled
at random with exactly $\rho_0L_xL_y$ particles, being $\rho_0=1/2$ the 
density of up spins. However, the number of particles on each row parallel to the field axis is not the same for all rows. This generates tiny density fluctuations along this direction, which are of the order of $(1/L_{x})^{-\frac{1}{2}}$, in agreement with the central limit theorem. According to equation (\ref{2op}), these fluctuations add up, and the amplitude of $OP$ depends on $L_x^{-1/2}$\footnote{These fluctuations are equivalent to the initial
  magnetization $m_0$ in the original
formulation of short time dynamics \cite{zhe}.}.
We have to
take into account this expression for the final form of the time evolution of $OP$. Setting
$b\simeq t^{1/z}$, eq. (\ref{2op}) becomes:
\begin{equation}
OP(t, \phi, L_x,L_y)=t^{-\beta/\nu_{\parallel}z} OP^{*}(1,t^{1/\nu_{\parallel}z}\phi, t^{-1/z}L_{x},t^{-\nu_{\perp}/\nu_{\parallel}z}L_y ).
\label{preop}
\end{equation}
\noindent Then, if $t^{-1/z}L_{x}$ is extracted out of the scale function in Eq. (\ref{preop}), we have:
\begin{equation}
OP(t, \phi, L_x,L_y)=t^{-\beta/\nu_{\parallel}z} (
t^{-1/z}L_{x})^{x}OP^{**}(1,t^{1/\nu_{\parallel}z}\phi,t^{-\nu_{\perp}/\nu_{\parallel}z}L_y),
\end{equation}
\noindent but since  $OP \simeq 1/L_{x}^{\frac{1}{2}}$, then  $x=-1/2$, so, the final expression for $OP(t)$ is the following:
\begin{equation}
OP(t,\phi,L_x)= L_x^{-1/2}t^{c_{1}}
OP^{**}(t^{1/\nu_{\parallel}z}\phi)\qquad L_x,L_y\rightarrow \infty, \qquad
\label{fdc}
\end{equation}
with
$c_{1}=(1-2\beta/\nu_{\parallel})/2z$ \cite{alsa}. \\

Furthermore, it is easy to show that the logarithmic derivative of
$OP$ with respect to $\phi$, given by Eq. (\ref{fdc}) at
criticality, behaves as

\begin{equation}
\partial_{\phi} ln  OP(t,\phi) \propto t^{c_{2}} ,  \qquad \\
\label{derfdc}
\end{equation}

\noindent where the exponent is $c_{2} = 1/\nu_{\parallel}z$. \\
On the other hand, starting
the system from the GSC configuration described above, and according
to
 the  scaling behavior proposed in \cite{alsa}, we have
\begin{equation}
OP(t,\phi,L_x,L_y)=b^{-\beta/\nu_{\perp}}
 OP^{***}(b^{-z}t,b^{1/\nu_{\perp}}\phi, b^{-1}L_{x},b^{-\nu_{\parallel}/\nu_{\perp}}L_{y}) ,
\label{gsc}
\end{equation}
 where $OP^{***}$ is another scaling function . Here we have also included the shape scaling factor $b^{-\nu_{\perp}/\nu_{\parallel}}L_y$. Proceeding in the same way as in the above case, we have, taking $b\simeq t^{1/z}$ in (\ref{gsc}) at
criticality, the following expression for $OP$:

\begin{equation}
OP(t)\propto t^{-c_{3}}\qquad L_x,L_y\rightarrow \infty, \qquad
\label{gsc1}
\end{equation}
 with an exponent $c_{3} = \beta/\nu_{\perp}z$.
Moreover, the derivative of Eq. (\ref{gsc1}) with respect to $\phi$ at criticality is given by

\begin{equation}
\partial_{\phi} OP(t) \propto t^{c_4} , \qquad
\label{derigsc1}
\end{equation}

\noindent where the exponent is $c_{4} = (1 -\beta)/\nu_{\perp}z$.\\

\section{Simulation  Results}
\label{4}

\subsection{Critical Temperature}

In this work we used rectangular and square lattices of different sizes
$L_{x},L_{y}$, in the range $128\leq L_x, L_y\leq 10000$ lattice units.  The critical dynamics of the model was investigated as a function of the shear magnitude, in the range $1/32\leq\dot{\gamma}\leq50$. The temperatures were measured in units of $J/k_{B}$, $k_B$ being the Boltzmann constant, and the time is measured in Monte Carlo steps (mcs), where one unit consists of $L_x L_y$ attempts for spin updates. The time evolution was sampled from 100 to 1000 realizations of the system, according to each initial condition and temperature.\\
 We begin showing our results by considering the critical dynamic evolution of the system
 when it is started from FDC configurations, and then coupled with  a thermal bath at $T\simeq T_c$. In figure \ref{fdcgd5},  the time evolution of $OP(t)$ is shown for a system where a shear field is applied with shear rate $\dot{\gamma}=5$. The best power law behavior is obtained at $T=T_c=2.660$ before $OP(t)$ reaches a saturation value due to finite size effects. On the other hand, if $T<T_c$ or $T>T_c$, $OP$ deviates from the power law behavior as equation (\ref{fdc}) states, and the curve shows an upward or
downward bending, respectively (see the curves corresponding to $T=2.655$ and $T=2.665$ in the same figure) 
\begin{figure}[H]
\centering
\includegraphics[height=15cm,width=11cm,clip=,angle=270]{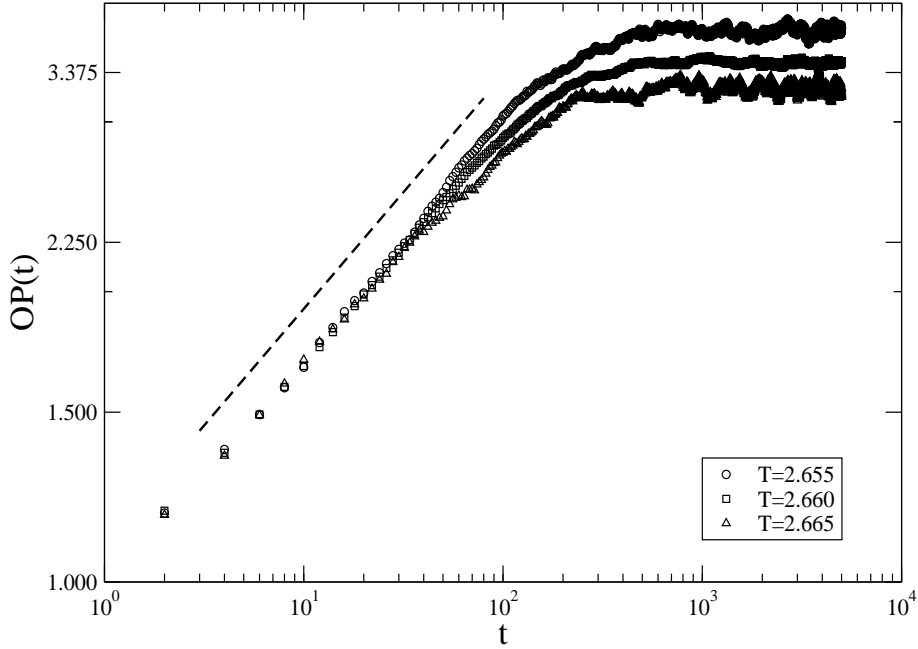}
%\includegraphics[height=15cm,width=11cm,clip=,angle=270]{op_fdc_allgds2b.ps}
%\centering\epsfig{figure=dibu.eps,width=\linewidth}
\caption{Log-log plot of the short time evolution of $OP(t)$
starting from FDC configurations, in a system with $L_x=L_y=512$  and
$\dot{\gamma}=5$. The best fit of the raw data gives a power law
behavior at $T_c=2.660$, as it is indicated by the dashed straight
line. Upward and downwards deviations from this behavior can also be
observed for $T=2.655$ and $T=2.665$, respectively.}
\label{fdcgd5}
%\end{center}
\end{figure}
The critical dynamic evolution was investigated by performing simulations also on rectangular lattices. The plots in figures \ref{fdcs} display the dynamic evolutions of $OP(t)$ at $T=T_c$ for two shear field magnitudes, $\dot{\gamma}=1/2$ and $\dot{\gamma}=10$ in a) and b) panels, respectively. The lattice sizes used are indicated in the legend of each plot by the notation $L_x$ $\times$ $L_y$. In the main plots of each figure, all early-time evolution exhibits the same power-law behavior with similar values of the exponent $c_1$ (equation (\ref{fdc})). Then, a saturation value is reached, $OP_{sat}$, that depends only on $L_x$, as it can be observed in lattices with longitudinal sizes $L_x=500$ and $L_x=1000$ for $\dot{\gamma}=1/2$ and $L_x=500$ for $\dot{\gamma}=10$ respectively. So, these plots show that the early-time critical evolution of the system is free from lattice shape effects \cite{kls}, because the relation between the longitudinal and transversal sizes is different for each studied case. The same was observed in the short-time critical evolution of the DLG model \cite{alsa}. In addition, the power law behaviors can be collapsed by rescaling $OP(t)$ by $L_{x}^{1/2}$ as it is proposed in equation (\ref{fdc}). This is shown in the insets of both figures
\begin{figure}[H]
\centering
\includegraphics[height=9cm,width=7cm,clip=,angle=-90]{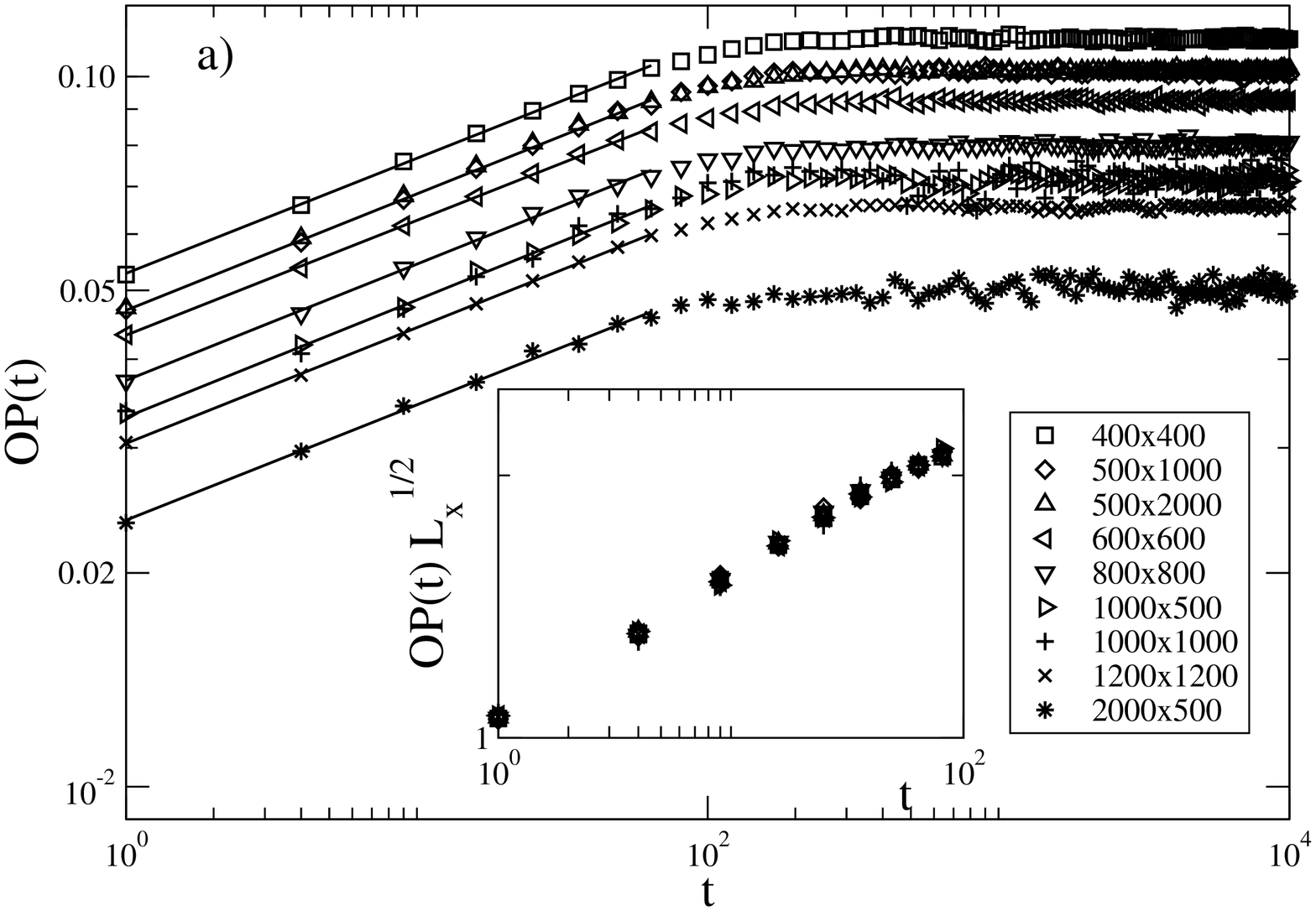}
\includegraphics[height=9cm,width=7cm,clip=,angle=-90]{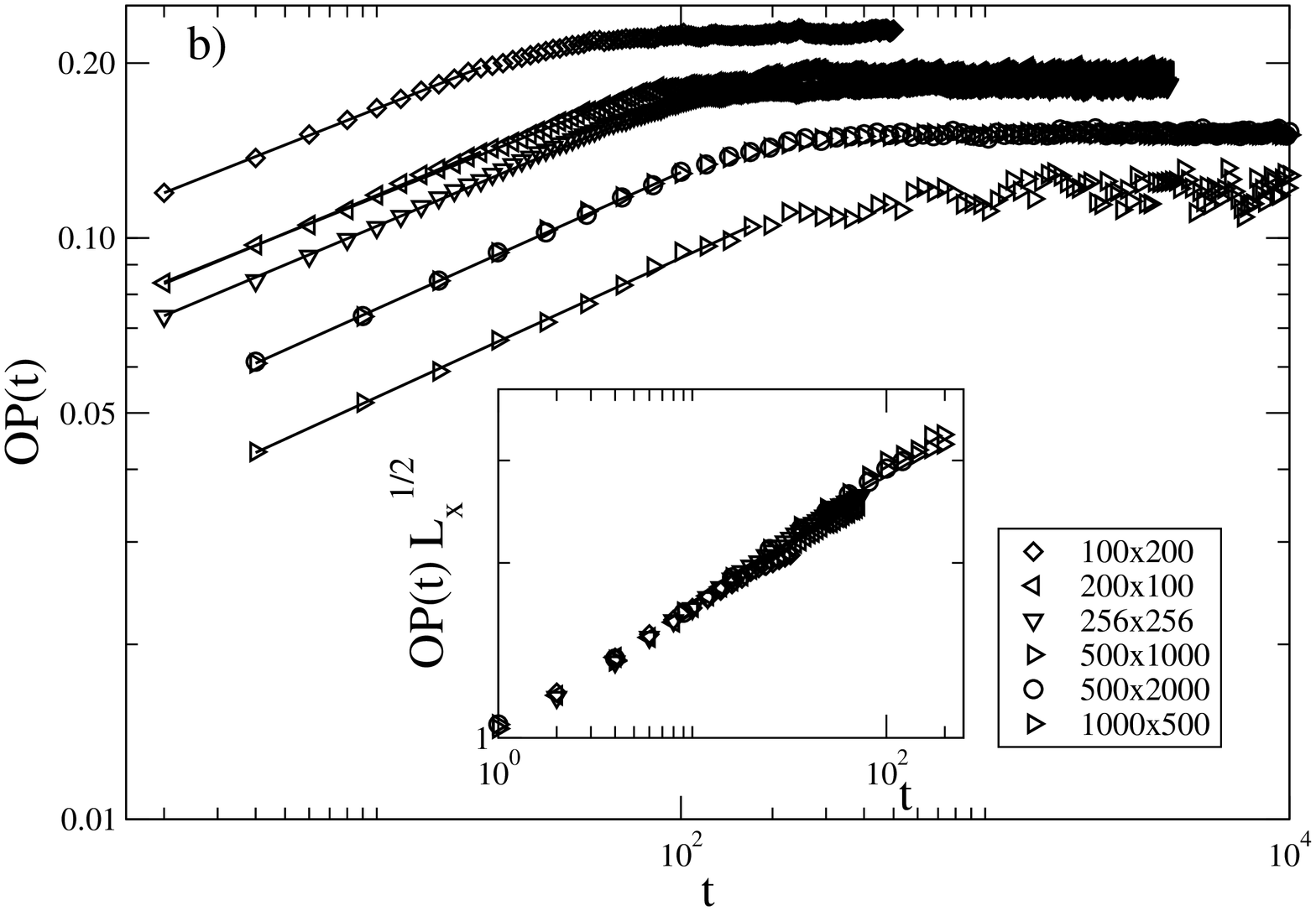}
\caption{Log-log plots of the time evolutions of $OP(t)$ at $T=T_c$, for the system in
 rectangular and square lattices of sizes $L_x \times L_y$, as indicated in the legends.
 In a) the external field has magnitude $\dot{\gamma}=1/2$ while in b) the shear
 field magnitude is $\dot{\gamma}=10$. The straight lines are least-square fits of the data.
 The insets in each figure show the collapse of the power law behaviors when  $OP(t)$ is multiplied
 by $ L_x^{1/2}$
 }
\label{fdcs}
\end{figure}

Then, the critical points of the system at several values of the
shear field magnitudes  $\dot{\gamma}$ were found. Figure
\ref{fdcgds} shows that the power-law behavior of $OP$ collapses for
large values of $\dot{\gamma}$, i.e. $\dot{\gamma}=5, 10$ and $50$,
and occurs at approximately the same critical temperatures for each
shear rate. These temperatures are larger than the estimated for the
equilibrium Ising model, $T_c(\dot{\gamma}=0)=2.269$ (see table 1).
However, the time evolution of $OP$ depends on the shear field if
$\dot{\gamma}$ is small, as it can be seen for the cases
$\dot{\gamma}=1/32$ and $\dot{\gamma}=1/10$. Although these
magnitudes are quite small, they are enough to raise the critical
temperature to $T_c(\dot{\gamma}=1/32)=2.29$ and
$T_c(\dot{\gamma}=1/10)=2.395$, both close but  greater than  the
critical temperature of the Ising model. This confirms that the
critical temperature depends on the shear field magnitude, as found
by theoretical studies \cite{gonepeli}.
\begin{figure}[H]
\centering
\includegraphics[height=15cm,width=11cm,clip=,angle=-90]{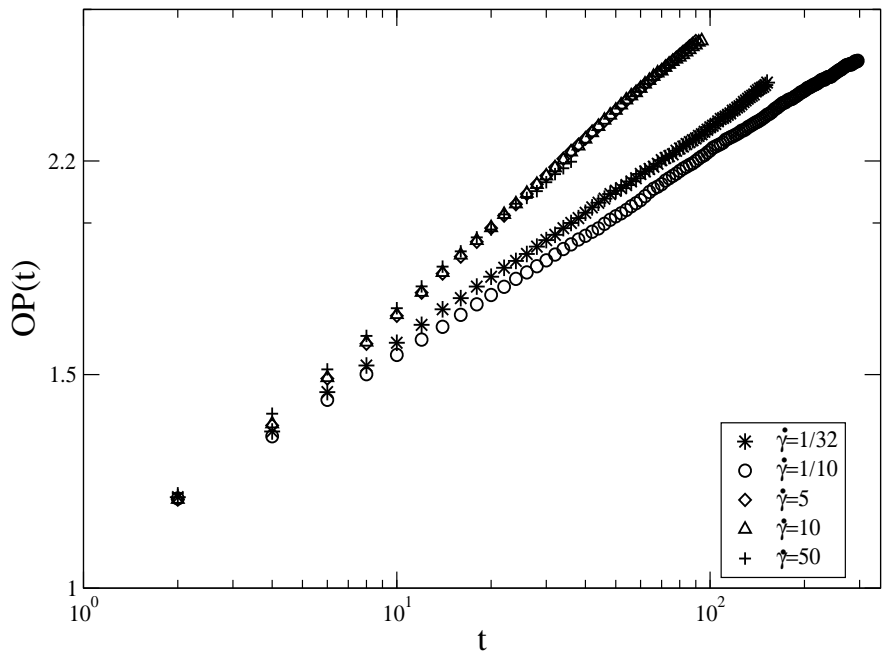}
\caption{Log-log plots of the time evolutions of $OP(t)$ at $T=T_c$
 corresponding  to systems with different shear fields.
The magnitudes $\dot{\gamma}$' are indicated in the legend.}
\label{fdcgds}
\end{figure}

Once that the critical temperatures $T_c$ corresponding to the
different  $\dot{\gamma}$'s  were collected, a diagram of critical
temperatures  versus $\dot{\gamma}$ can be performed. Figure
\ref{diag_fase} shows that two regimes can be distinguished. In the
first regime, the critical temperature grows with $\dot{\gamma}$ as
a power law, i. e. $T_c(\dot{\gamma})/T_c(0)-1$ $\propto$
$\dot{\gamma}^\psi$.
 The value of the exponent
was estimated in $\psi=0.52(3)$, which is consistent with that calculated theoretically in
\cite{gonepeli}. In this work (ref. \cite{gonepeli}), the critical transition in the 
 $\varphi^3 \rightarrow \langle\varphi^2\rangle\varphi$ approximation was studied in a scalar field model based on a convection-diffusion equation with Landau-Ginzburg free energy, with the average $\langle\varphi^{2}\rangle$ self-consistently determined. Above the lower critical dimension, the exponent $\psi$ was evaluated to be $1/2$ and $1/4$ for the cases with non-conserved and conserved order parameter respectively.

Then, $T_c$ crosses over to a saturation regime at larger $\dot{\gamma}$'s,
saturating at $T_c(\dot{\gamma})\backsim 1.18 T_c(0)$, where $T_c(0)\approxeq2.269$ is the
critical temperature of the 2D Ising model. The increase of the critical temperature by action
of the external field, and a posterior saturation regime was also  observed in the  DLG model \cite{kls}. We want to remark also that in real fluids there is also a negative contribution to the shift of the critical temperature coming from hydrodynamics interactions \cite{ok}. The total shift of $T_c$ results to be negative for fluids with low molecular weight \cite{ok}, was also found in experiments \cite{bg}. A review for other systems is given in \cite{onuki}.

\begin{figure}[H]
\centering
\includegraphics[height=15cm,width=11cm,clip=,angle=270]{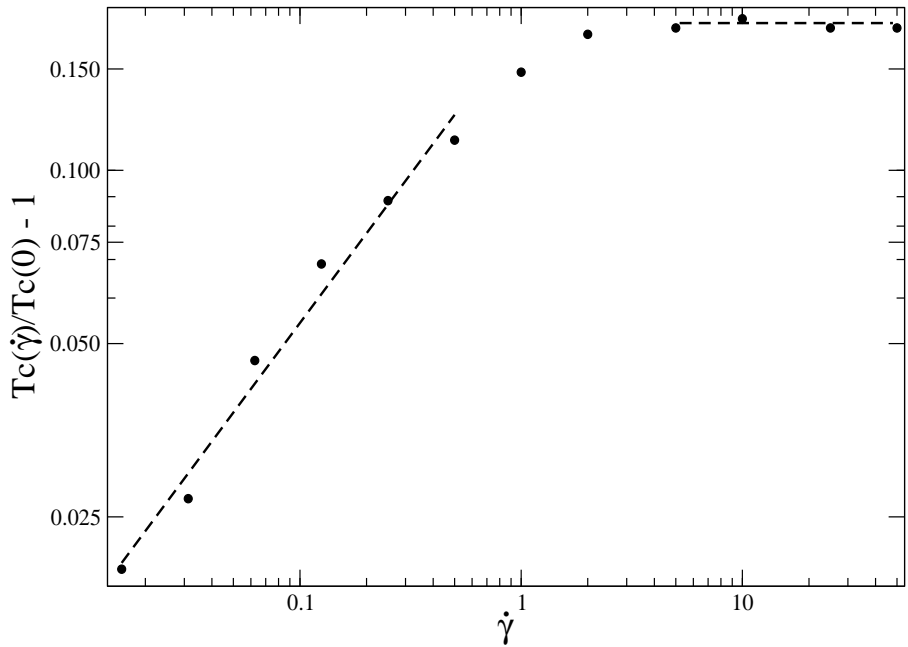}
%\includegraphics[height=15cm,width=11cm,clip=,angle=270]{DTc_vs_gd_semII.ps}
%\centering\epsfig{figure=dibu.eps,width=\linewidth}
\caption{Diagram of critical temperatures
$T_c(\dot{\gamma})/T_c(0)-1$ versus $\dot{\gamma}$, in log-log
scale. The fit of the data points in the growing regime gives a
slope $\psi=0.52(3)$.}\label{diag_fase}
%\end{center}
\end{figure}

On the other hand, if the system is started from the GSC initial condition (i.e. magnetization equal to 1),
and then it is left to evolve at the working temperature $T\simeq T_c$, $OP(t)$ decreases, and
 follows a power law behavior at $T=T_c$. Upwards or downwards deviations are observed
  according if $T<T_c$ or $T>T_c$, respectively. Figure \ref{gscgd5} exhibits the evolution of
  $OP(t)$ for the same parameters of  figure \ref{fdcgd5}. A clear power law behavior can be observed at $T=2.66$, which is exactly the same temperature found when the system was started from FDC configurations. This is precisely the signature of a second-order phase transition in the model. In addition, the expected deviations from the power law behavior at $T=2.65$ and $T=2.67$ are also observed.

\begin{figure}[H]
\centering
\includegraphics[height=15cm,width=11cm,clip=,angle=270]{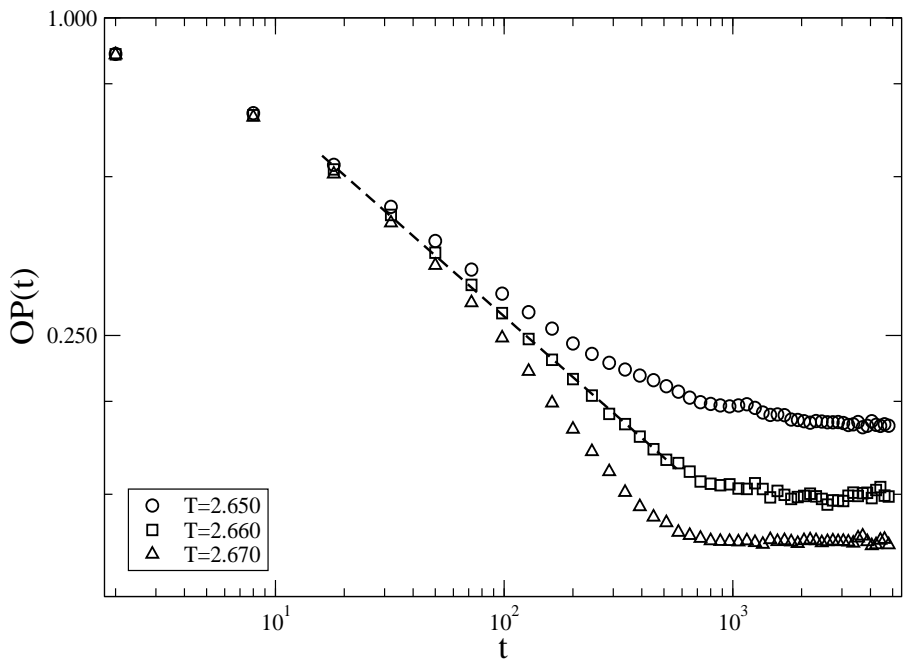}
%\includegraphics[height=15cm,width=11cm,clip=,angle=270]{gsc_gds_all_3b.ps}
%\centering\epsfig{figure=dibu.eps,width=\linewidth}
\caption{Log-log plot of the time evolutions of $OP(t)$
corresponding to a system set in a square lattice of size $L_x=L_y=512$
with shear rate $\dot{\gamma}=5$. The  working temperatures of the
thermal bath are indicated in the legend. A power law behavior is
observed for $T_c=2.660$ $J/k_B$. The dashed line is a fit of the
numerical data.}\label{gscgd5}
%\end{center}
\end{figure}

Proceeding in the same way we did for the
evolutions started from FDC's,  the critical behavior
of the system was investigated when it is started from GSC configurations in
rectangular lattices. Figures \ref{gscls} a) and b) show the critical
time evolution of $OP$ when the system is initiated from a
GSC configuration, corresponding to $\dot{\gamma}=1/2$ and
$\dot{\gamma}=10$, respectively. It is important to remark that the best power law behavior was obtained when the systems evolved at the same critical temperatures found when they were initiated from FDC configurations. According to the results,
the transversal size $L_y$ does not play a relevant role in the
critical behavior of the system  as it is shown in the evolutions in
lattices with $L_x=500$ (figure \ref{gscls} a) and b)) and also with
$L_x=2000$ (figure \ref{gscls} b)), but rather the longitudinal size
$L_x$ is relevant. This is in agreement with the results previously
exhibited in figure \ref{fdcs}, and allow to conclude that that the
critical evolution of the system is independent of lattice shape
effects for both initial configurations.

\begin{figure}[H]
\centering
\includegraphics[height=9cm,width=7cm,clip=,angle=-90]{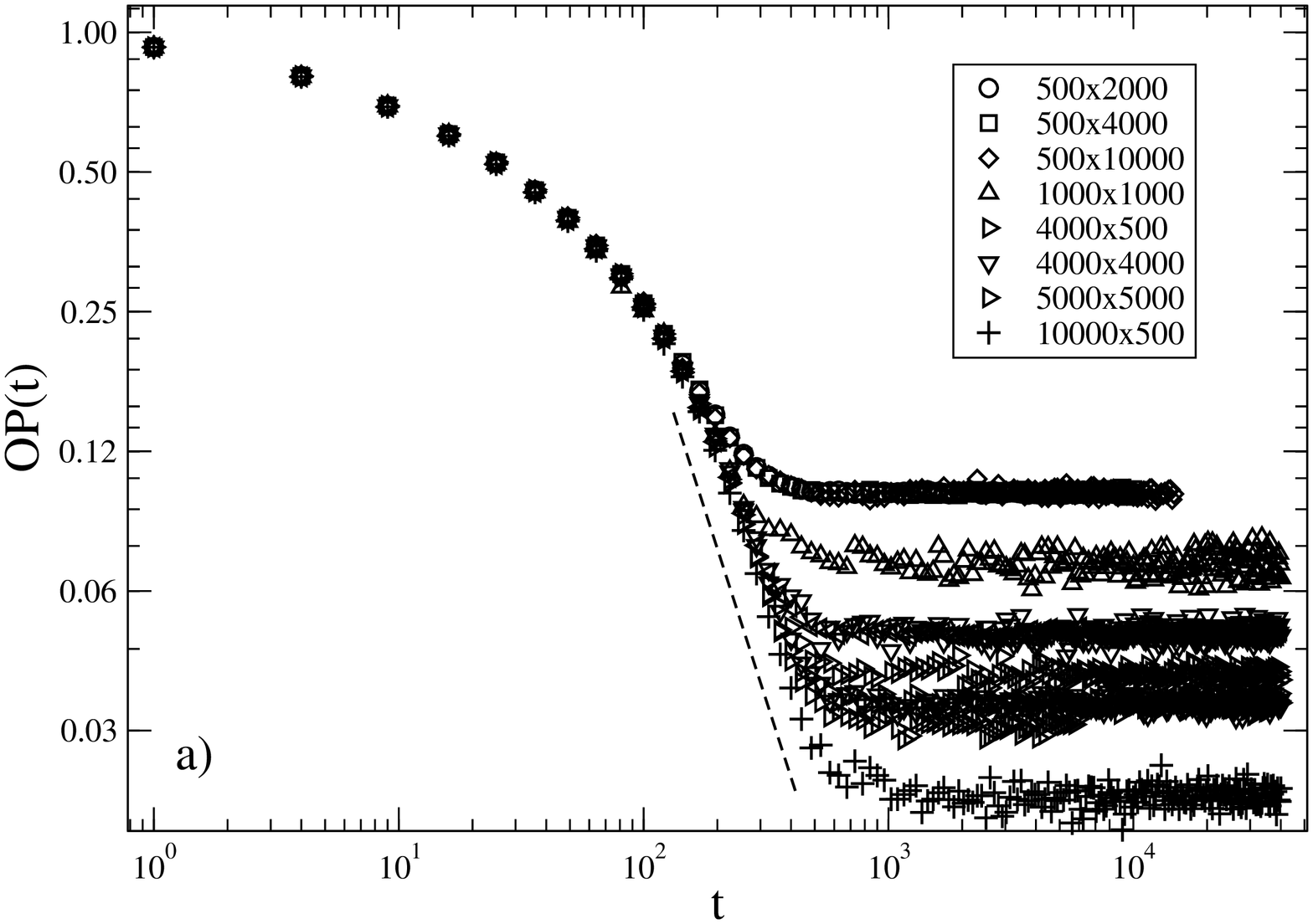}
\includegraphics[height=9cm,width=7cm,clip=,angle=-90]{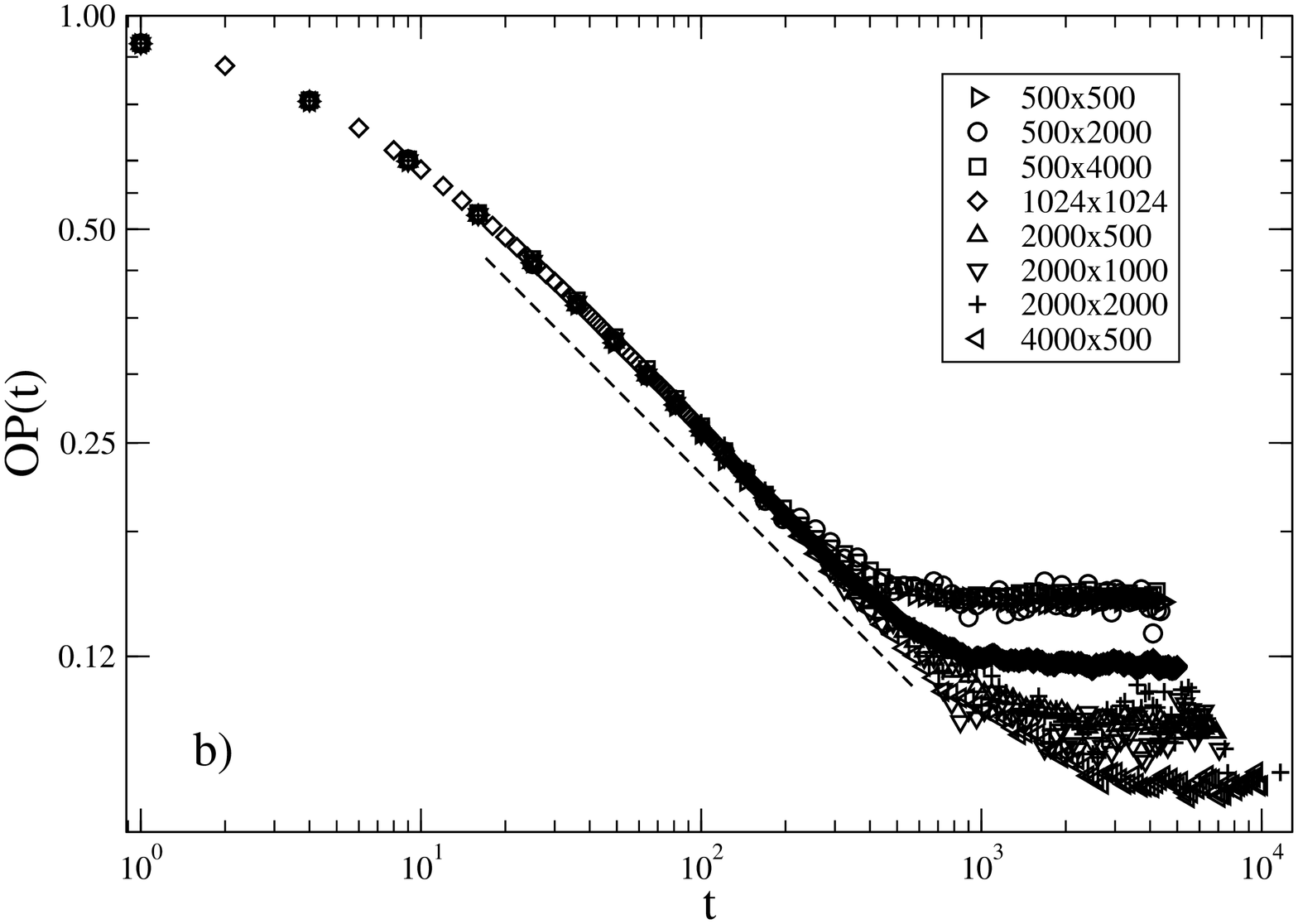}
\caption{Log-log plots of the time evolutions of $OP(t)$ at $T=T_c$, for the system set in
 rectangular and square lattices of sizes $L_x \times L_y$, as
 indicated in the respective legends. In a) the external field has magnitude $\dot{\gamma}=1/2$
 while in b) the shear field magnitude is $\dot{\gamma}=10$. The dashed lines  displaced are
  least-square fits of the obtained data.}
\label{gscls}
\end{figure}

\subsection{Critical Exponents}

 Focusing our attention on the power law behavior at the $T=T_c$, the exponents  $c_1$ and $c_2$  corresponding
to the dynamic evolution of $OP$ (equation (\ref{fdc})) and its
logarithmic derivative $\partial_{\tau} ln OP(t,\tau)$ (equation
(\ref{derfdc})) can be estimated. Table \ref{tablac}  enlists the
obtained values of $T_c$ and the mentioned exponents for small and
large values of $\dot{\gamma}$. The values of $c_1$ and $c_2$, for
the case of $\dot{\gamma}=1/32 $ are different compared with the
those at larger  $\dot{\gamma}$, as it was  already observed in  figure  \ref{fdcgds}.  Furthermore, the
value of $c_1$ for $\dot{\gamma}=50$ is also slightly different from
the corresponding cases of $\dot{\gamma}=5$ and $\dot{\gamma}=10$.
This seems to be a rare behavior, since there exists a saturation
regime for the critical temperature at these values of
$\dot{\gamma}$'s (figure \ref{diag_fase}), and we are induced to
think that the dynamic behavior of the system is independent of the
field magnitude in this limit. Opposite to that, the estimated
values of $c_2$ are quite similar in the limit of large
$\dot{\gamma}$'s.

\begin{table}[H]
\centering
\begin{tabular}{|c|c|c|c|}
  \hline
  $\dot{\gamma}$ & $T_c$ &$c_1$&$c_2$ \\
  \hline
  \hline
  1/32& 2.29&0.180(8)&0.99(2) \\
%\hline
 % 1/10& 2.395&0.167(3)&0.91(1) \\
   \hline
  5 &2.66&0.239(1)&0.84(1)\\
   \hline
   10 &2.675&0.238(1)&0.87(1)\\
   \hline
  50 &2.675&0.224(1)&0.85(1) \\
   \hline
 % DLG (E=50) &3.20(1)&0.114(5)&0.406(10)\\
   %\hline
  \hline
\end{tabular}
\caption{Exponents $c_1$ and $c_2$, obtained from FDC initial conditions, corresponding to the values of $\dot{\gamma}$ enlisted in the first column.
%, and the values of the DLG model at large values of the driving field (taken from ref. \cite{alsa}).
}\label{tablac}
\end{table}

\noindent Table \ref{tablac2} enlists the values of the
exponents $c_3$ and $c_4$  that were obtained from a least-square fits of the critical evolution of the system when it is initiated from the GSC configurations (see equations (\ref{gsc1}) and (\ref{derigsc1})).  Here, the same situation that happened for $c_1$ is found for $c_3$.
In fact, the value of $c_3$ for $\dot{\gamma}=1/32$  is different form the rest of
the corresponding values at larger $\dot{\gamma}$'s,
 and the value of $c_3$ for $\dot{\gamma}=50$ is also different from the values
 estimated for $\dot{\gamma}=5$ and $\dot{\gamma}=10$. On the other hand, the values
 of $c_4$ are similar for all the reported cases of $\dot{\gamma}$'s.

\begin{table}[H]
\centering
\begin{tabular}{|c|c|c|c|}
  \hline
  $\dot{\gamma}$ & $T_c$ &$c_3$&$c_4$ \\
  \hline
  \hline
  1/32& 2.29&0.076(1)&0.65(2) \\
\hline
%  1/10& 2.395&1.030(1)&0.66(1) \\
 %  \hline
  5 &2.66&0.401(1)&0.63(1)\\
   \hline
   10 &2.675&0.405(5)&0.62(1)\\
   \hline
  50 &2.675&0.360(7)&0.62(5) \\
   \hline
%  DLG (E=50)&3.20(1)&0.254(15)&0.516(15)\\
%   \hline
  \hline
\end{tabular}
\caption{Exponents $c_3$ and $c_4$ obtained from the GSC configuration, corresponding to the values of $\dot{\gamma}$ enlisted in the first column.
%, and the values of the DLG model at large values of the driving field (taken from ref. \cite{alsa}).
}\label{tablac2}
\end{table}
\noindent According to the equations developed in section \ref{3}, the critical exponents of the second-order phase transition, are obtained by combining the estimated exponents $c_1$, $c_2$, $c_3$, and $c_4$ enlisted above, corresponding to each case of $\dot{\gamma}$ investigated. Table \ref{tablacc}  resumes the obtained results and includes  the critical exponents of both the Ising and DLG model for the sake of comparison. Since this issue for the case of the DLG model still remains an open problem, the obtained theoretical values from all proposed theories exposed in refs. \cite{kls} and \cite{gallegos1} are included.

\begin{table}[h]
\centering
\begin{tabular}{|c|c|c|c|c|}

 \hline
  $\dot{\gamma}$ & $\beta$&$z$&$\nu_{\perp}$&$\nu_\parallel$ \\
  \hline
  \hline
  1/32& 0.105(1)&1.77(7)&0.78(1)&0.57(5) \\
%  \hline
 % 1/10& 0.61(1)&0.70(1)&0.85(2)&1.60(8) \\
  \hline
  5 &0.39(1)&0.88(1)&1.10(1)&1.35(2)\\
  \hline
  10&0.39(1)&0.86(1)&1.14(1)&1.34(1)\\
  \hline
  50 &0.37(1)&0.94(1)&1.10(1)&1.30(6)\\
  \hline
  Ising 2d &0.125 &2.16&1&1\\
  \hline
  DLG(E=50) (ref. \cite{kls}) &1/2&$\approx$4/3&$\approx$3/2&$\approx$1/2\\
  \hline
  DLG (E=50) (ref. \cite{gallegos1}) &$\approx$0.33&$\approx$1.998&$\approx$1.22&$\approx$0.63\\   \hline
  \hline
\end{tabular}
\caption{Table of the calculated critical exponents for each case of
$\dot{\gamma}$ employed. The critical exponents corresponding to both theories of the critical phase transition of the Ising and the DLG models are also included for comparison. Since there is no anisotropy in the Ising model $\nu_\bot=\nu_\parallel=\nu$ must be read. The results of this table do not fit with the relation $\psi=1/\nu z$ suggested in \cite{ok}. See, however, in Sect. 5, the discussion about our results our small shear rates.}\label{tablacc}
\end{table}

\noindent An overview of this table deserves some comments.  First, the
calculated value of $\beta$ at  $\dot{\gamma}=1/32$ is close to
$\beta=1/8$ calculated for the 2d Ising model. This similarity could
drive us to think that the effects of the shear field are
negligible, but this is not the case, as it is evidenced by the
values of the rest of the critical exponents. In fact, anisotropy
effects are important, even if a small external field,
in this case $\dot{\gamma}=1/32$, is applied. The dynamic exponent $z (\dot{\gamma}=1/32)=1.77(7)$
indicates that the correlation length (in the longitudinal direction, see  \ref{lcorr})
grows faster with time than the corresponding one in the Ising model $z=2.16$, and the difference between
$\nu_\bot(\dot{\gamma}=1/32)=0.78(1)$ and $\nu_\parallel(\dot{\gamma}=1/32)=0.57(5)$ reveals an anisotropic
 critical behavior even at small shear rate values. \\
The situation is different for the critical exponents at the largest values of $\dot{\gamma}$ investigated.
The values of the  exponents are similar between each other,  suggesting that the critical behavior does not depend of the applied field. This fact is also present in figure \ref{diag_fase},
 where the critical temperature is approximately the same for the largest values of $\dot{\gamma}$ used.
 Furthermore, we also noticed that $\nu_\bot>\nu_\parallel$ for $\dot{\gamma}=1/32$, while it happens
 the opposite at large $\dot{\gamma}$.  At the moment, a reasonable explanation for this issue is not possible due to the lack of a theoretical framework about the critical behavior of this model.\\
To end this section, one final comment is appropriate. In view of
the values exposed in table 3, the computed critical exponents do
not belong to the universality classes of the Ising or of the DLG
models respectively. This fact is not surprising since, as we have
seen, the shear rate affects the critical behavior of the model by
inducing anisotropic effects in the equilibrium model that changes
its behavior. The case of the DLG model is different. Although both
models have a similar phase behavior, the particle dynamics defined
for the DLG model conserves the number of particles while our model
does not.  This difference will probably affect the values of the
critical exponents, and in consequence there is no reason to expect
that both models will belong to the same universality class.

\subsection{Longitudinal Correlation}
\label{lcorr}

In section \ref{3}, it is assumed that the dynamic increase of the  longitudinal correlation length $\xi_{\parallel}$ and the breakage of the corresponding transversal one $\xi_{\perp}$ at $T=T_c$ are due to the anisotropy effects induced by the external shear field. This will cause that the short-time critical dynamic evolution of the system initiated from either the FDC or GSC configurations will depend only on the dynamic critical exponent $z_{\parallel}=z$ that describes the critical dynamic increase of $\xi_{\parallel}$. \\
To show our hypothesis, we performed a scaling of the
whole curve of the $OP(t)$ for the cases of $\dot{\gamma}=1/2$
(small driving field) and $\dot{\gamma}=10$ (large driving field).
We propose a phenomenological scaling in the spirit of the scaling
form used by Family and Vicsek to describe  the roughness growth of
interfaces \cite{famvis} (obviously in a different context not related to ours).
This is given by the following expression
\begin{equation}
OP=L_x^{-\omega_{i}}f(t/L_x^{z}), \label{f-v}
\end{equation}

\noindent where $L_x$ is the longitudinal size, according to the
results exhibited in figures \ref{fdcs} and \ref{gscls} a) and b),
respectively. The exponent $\omega_{i}$, $i=$FDC or GSC, is the
exponent that has into account the finite-size critical behavior of
$OP$, and $z$ is the relevant dynamic critical exponent. The idea of
this scaling form is simple: if  all the curves can be collapsed by
using the same dynamic exponent $z$  independently of the initial
condition used to start the simulations, it can be demonstrated
numerically that only one correlation is relevant in the critical
dynamic behavior. Furthermore, equation (\ref{f-v}) must contain
both the critical dynamic behavior of $OP$ according to equations
(\ref{preop}) and (\ref{gsc1}) at the early times of evolution, and
the
 finite size behavior in the limit of large times ($t\rightarrow\infty$)
 where the correlation length is comparable to $L_x$. Therefore we have that  $f(u)$
 must be $f\propto (t/L_x^z)^{-\beta/\nu_{\parallel}z}$ or $f\propto (t/L_x^z)^{-\beta/\nu_{\perp}z}$ at early times $u\ll1$, depending if  the initial condition is $FDC$ or $GSC$ respectively.  This fixes the finite-size exponent $\omega_i$ in $\omega_{FDC}=\beta/\nu_{\parallel}$ or $\omega_{GSC}=\beta/\nu_{\perp}$ according to the initial configuration used to start the simulations.\\
Figures \ref{satgd05} and \ref{satgd10} a) and b) exhibit both the
finite-size dependence of $OP(t)$ with the longitudinal size $L_x$
(insets), and the scaling function $f(t/L_x^z)$ (main plots), for
the small and large external fields, represented by
$\dot{\gamma}=1/2$ (figures \ref{satgd05}) and $\dot{\gamma}=10$
(figures \ref{satgd10}) respectively. In all plots, the finite-size
dependence is obtained by calculating the saturated value of
$OP(t)$, $OP_{sat}$, from figures \ref{fdcs} and \ref{gscls}, which
were plotted versus $L_x$.

%\newpage
\begin{figure}[H]
\centering
\includegraphics[height=9cm,width=7cm,clip=,angle=270]{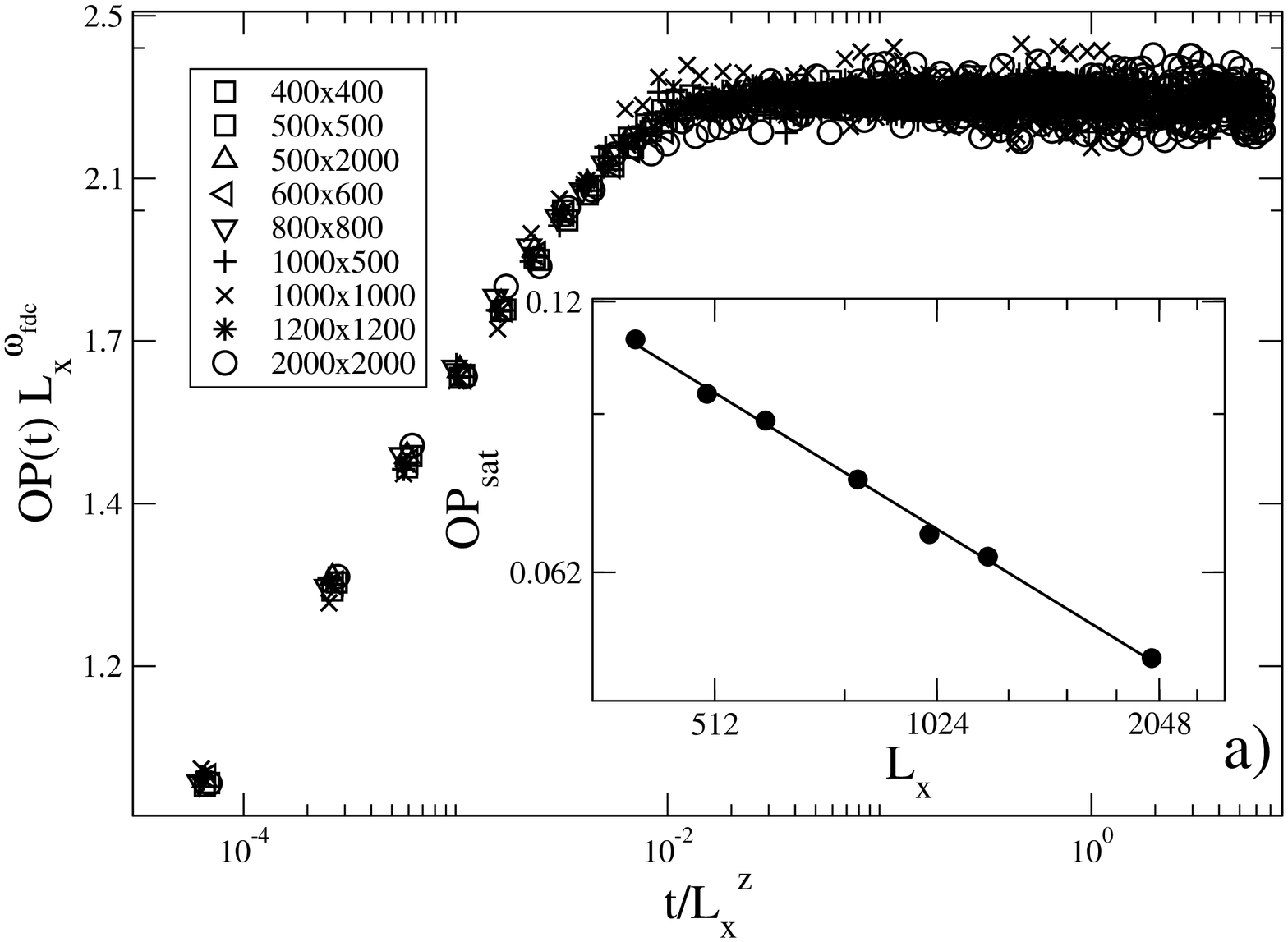}
\includegraphics[height=9cm,width=7cm,clip=,angle=270]{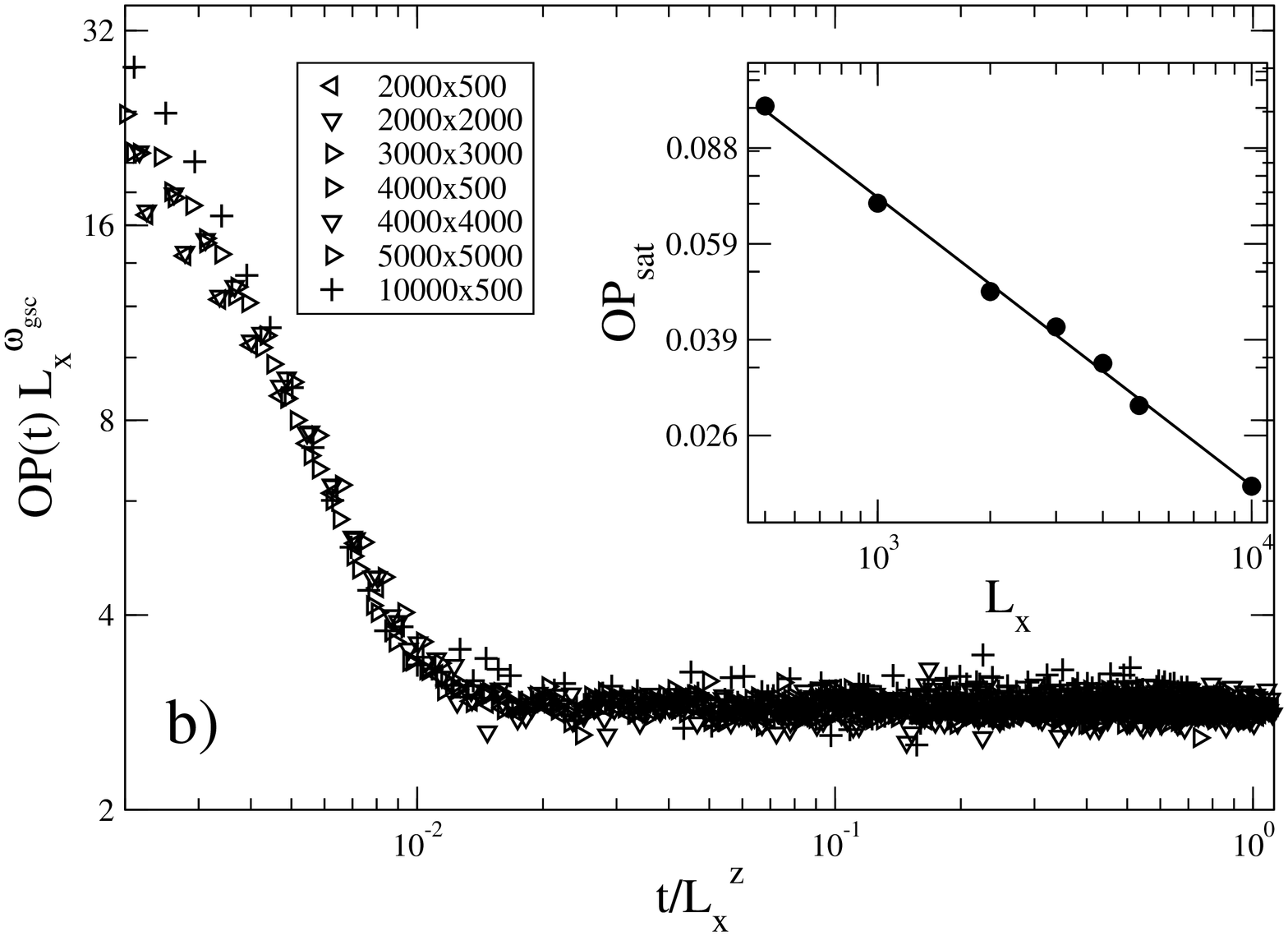}
%\includegraphics[height=15cm,width=11cm,clip=,angle=-90]{gd05_fdc_scal.ps}
%\centering\epsfig{figure=dibu.eps,width=\linewidth}
\caption{Log-log plots of the scaling functions $f(t/L^z)$ (see equation (\ref{f-v})) of the time evolution of $OP(t)$
 when the system is started from FDC and GSC configurations (main plot of figure a) and b) respectively) corresponding to a system with external field of magnitude $\dot{\gamma}=1/2$.
The insets of both figures show the finite-size dependence and the
corresponding power-law fit in a double logarithmic scale. The
values of the corresponding fits are $\omega_{FDC}=0.50(1)$ (inset
of a) ) and  $\omega_{GSC}=0.53(1)$ (inset of b)), respectively.
}\label{satgd05}
%\end{center}
\end{figure}
%\newpage

As it can be observed in the insets of the plots in figures \ref{satgd05} a) and b) the size dependence of $OP$ in the long time regime can be well fitted by a power law as it is proposed in equation (\ref{f-v}). The estimated exponents $\omega_{fdc}=0.50(1)$ and $\omega_{gsc}=0.53(1)$ were not consistent with the expected values $\omega_{FDC}=\beta/\nu_{\parallel}=0.263(3)$ or $\omega_{GSC}=\beta/\nu_{\perp}=0.242(2)$. However, the good collapses performed with the same $z(\dot{\gamma}=1/2)=1.45(3)$ exhibited in the main plots of  both figures clearly evidences that only longitudinal correlations are relevant in the critical evolution of the system at short times.\\

On the other hand, the insets of figure \ref{satgd10} a) and b), show that the finite-size behavior of $OP(t)$ is also a power law for the case of large shear field magnitudes, represented by $\dot{\gamma}=10$. Opposite to the case with $\dot{\gamma}=1/2$, the estimated values of  $\omega_{fdc}=0.29(2)$ and $\omega_{gsc}=0.34(3)$ were  in agreement with the expected values $\omega_{FDC}=\beta/\nu_{\parallel}=0.291(2)$ or $\omega_{GSC}=\beta/\nu_{\perp}=0.342(2)$ calculated from Table 3. The good collapses performed with $z(\dot{\gamma}=10)$ displayed in the main plots of the figures allow us to conclude that the same behavior observed for small $\dot{\gamma}'s$ is also exhibited by systems with large values of the external fields.\\

\begin{figure}[H]
\centering
\includegraphics[height=9cm,width=7cm,clip=,angle=270]{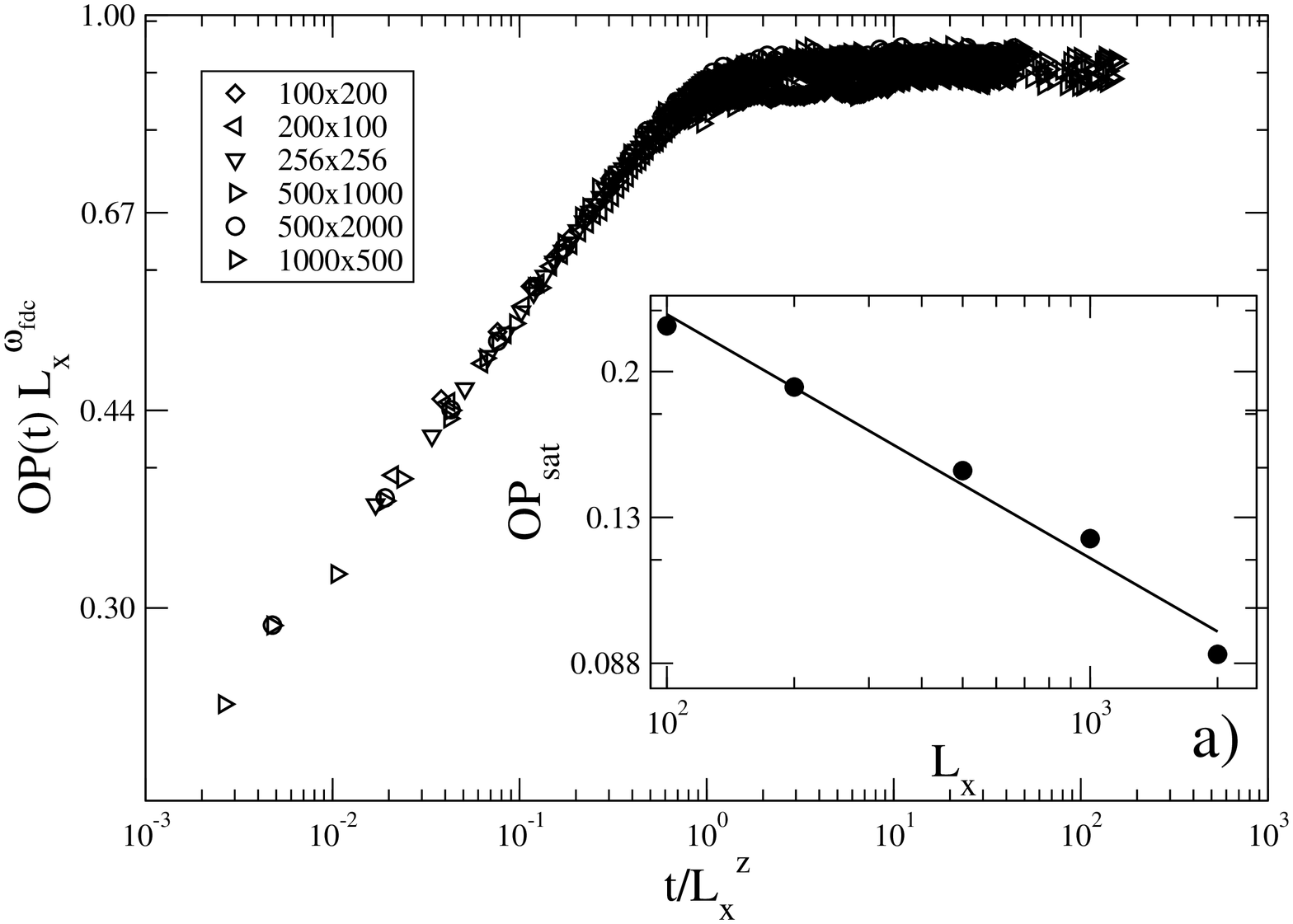}
\includegraphics[height=9cm,width=7cm,clip=,angle=270]{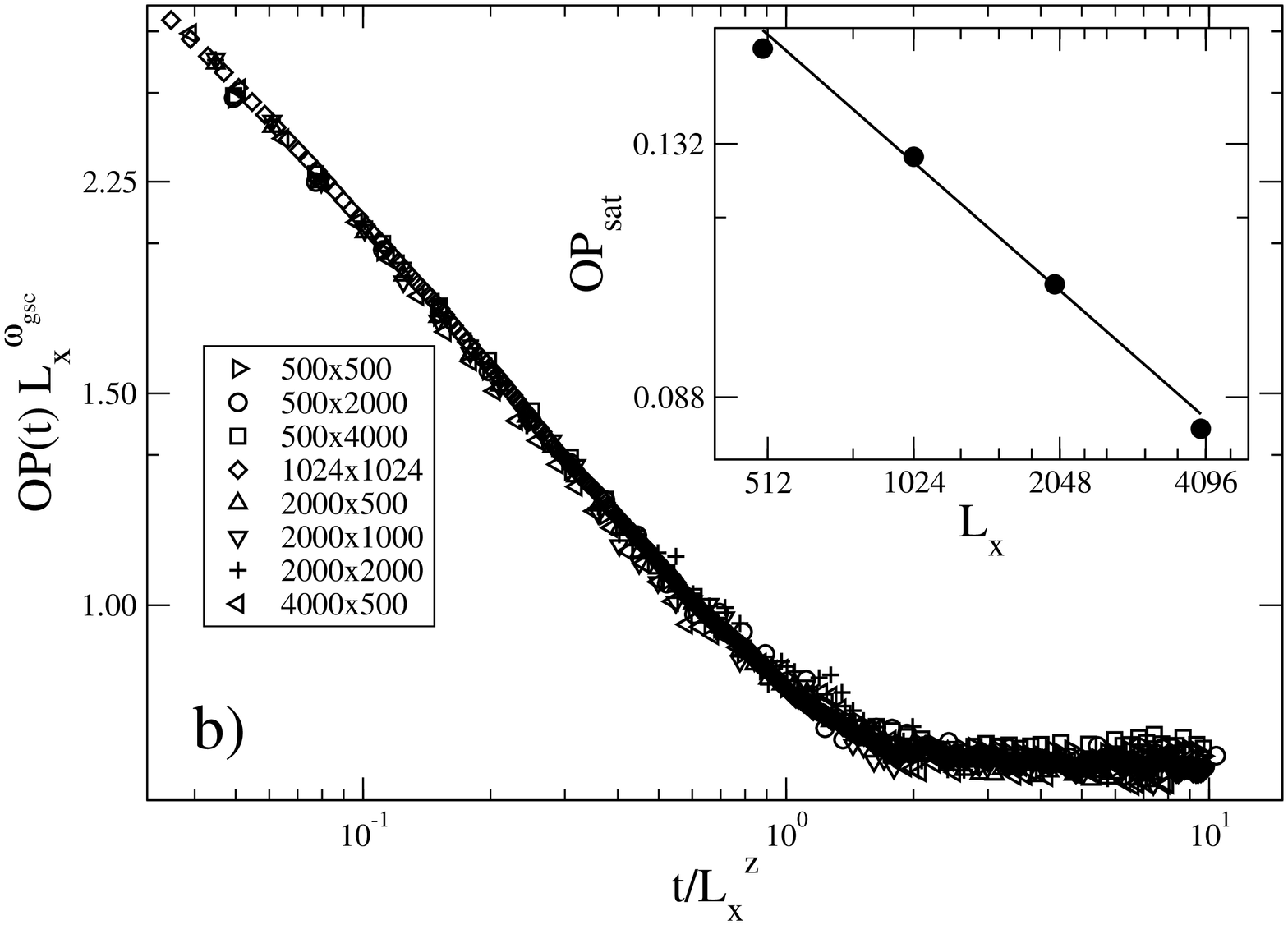}
%\includegraphics[height=15cm,width=11cm,clip=,angle=-90]{gsc_scal_gd10.ps}
%\centering\epsfig{figure=dibu.eps,width=\linewidth}
\caption{Log-log plots of the scaling functions $f(t/L^z)$ (see equation (\ref{f-v})) of the time evolution of $OP(t)$
 when the system is started from FDC's and GSC's configurations
 (main plot of figure a) and b) respectively) corresponding to a system with external field of magnitude $\dot{\gamma}=10$.
The insets of both figures show the finite-size dependence and the
corresponding power-law fit in a double logarithmic scale. The
values of the corresponding fits are $\omega_{fdc}=0.29(2)$ (inset
of a) ) and  $\omega_{gsc}=0.34(3)$ (inset of b)), respectively.}
\label{satgd10}
%\end{center}
\end{figure}

To summarize. we have shown that the critical dynamic evolution of
the system, started from either FDC's or GSC initial configurations,
can be scaled with the dynamic critical exponent $z$ proposed in
Section \ref{3}.  Based on the arguments exposed in Section \ref{3},
we can conclude that, in the short time limit of the critical
evolution, the correlations along the field axis (longitudinal) are
more relevant than transverse (perpendicular) correlations.
Furthermore, the obtained finite-size exponents $\omega_i$ are only
in accordance with those calculated using the critical exponents
enlisted in Table 3 corresponding to $\dot{\gamma}=10$, while at
$\dot{\gamma}=1/2$ they differ by a factor of nearly 2.

%\newpage

\section{Discussion and Conclusions}
\label{5}

In this work,  the second-order phase
transition in the 2d non-conserved Ising model under the action of an
external driving shear field was investigated by studying the critical evolution of the system in the short-time regime.
In order to apply this method, the dynamic evolution of the system at $T\simeq T_c(\dot{\gamma})$ was monitored when it is initiated from fully disordered initial configurations (FDC), and from the completely ordered configuration (GSC).

\noindent Starting the system from FDC's configurations,  the time evolution of the order
parameter $OP$ follows a power law behavior at the critical temperature $T_c$,
while at slightly  different values of $T$ the power law is modulated by a scaling function that
bends upwards or
 downwards depending  if the temperature is  less or greater than $T_c$,
respectively. The  critical evolution was studied on square and
rectangular lattices of different sizes $L_x$ and $L_y$, and  the
results  indicate that the short-time critical evolution is free of
shape effects. Furthermore, the saturation value reached by $OP$
depends only on $L_x$.
\noindent The critical evolution started from FDC's configurations was studied
for different values of  $\dot{\gamma}$, and the diagram of reduced
temperatures $T_c(\dot{\gamma})/T_c(0)-1$
 versus  $\dot{\gamma}$, $T_c(0)=2.269$ $J/k_B$ was drawn. As a first observation,
  all the values found for $T_c(\dot{\gamma})$
 are always greater than the 2d critical temperature of the Ising model,
 that is typical for models driven out of equilibrium by an external field.
Furthermore, two
 regimes can be distinguished: 1) a \textsl{growing regime} where
$T_c(\dot{\gamma})/T_c(0)-1$
  $\propto$ $\dot{\gamma}^\psi$. The exponent $\psi$ was calculated by means of
a linear regression fit,
  giving $\psi=0.52(3)$, which is consistent with theoretical predictions in ref.
\cite{gonepeli};
   2) a \textsl{saturation regime}, where  $T_c(\dot{\gamma})$ does not
change appreciably with
    $\dot{\gamma}$. In this regime  $T_c(\dot{\gamma})\sim 1.18 T_c(0)$.
A similar  diagram
    was already observed in the DLG model \cite{kls}, where
 the temperature grows with the magnitude of the driving field and then
saturate at large values.\\
On the other hand, the critical dynamic behavior of the model was
also investigated when it is initiated from the ground state
configuration (GSC).  A decreasing power law is observed for $OP(t)$
at the same $T_c(\dot{\gamma})$  found when the system was started
from FDC configurations. This evidences that the model experiences a
second order phase transition, as expected. Also in this case, the
critical evolution of the system was  simulated on rectangular and
square lattices of different sizes. It was found that the critical
evolution is independent of the shape of the lattice and the
saturation value of $OP$ only depends on the longitudinal size
$L_x$, as for evolutions initiated from FDC configurations. So, it
is concluded as a general result that in the short-time scale, the
system critical evolution is free of shape effects, as it is also
observed  for the critical evolution of the DLG model in the same time interval \cite{alsa}.

\noindent Then, the quantities  $c_i$ ($i=1,4$), defined in sect. 3, were
studied in order to calculate the critical exponents of the
transition. Starting from FDC's initial configurations, the dynamic
critical behavior at small $\dot{\gamma}$ is slower than  at larger
$\dot{\gamma}$'s. As it can be observed in figure \ref{fdcgd5} and
in Table 1, the order parameter exponent $c_1$ is smaller at
$\dot{\gamma}=1/32$ than
%the rest of the values reported
at larger $\dot{\gamma}$'s. Also, its logarithmic derivative
exponent $c_2$ is different for this case. On the other hand, the
values of $c_1$ and $c_2$ are more stable for larger
$\dot{\gamma}$'s, although $c_1$ at $\dot{\gamma}=50$ is slightly
smaller than the estimated values for $\dot{\gamma}=5, 10$ . A
similar scenario was found for the order parameter $c_3$ and its
derivative $c_4$ exponents starting from the GSC configuration. \\
By combining these exponents, the static and dynamic critical
exponents $\beta$, $\nu_{\bot}$, $\nu_{\parallel}$ and $z$ were
calculated and
 are enlisted in table 3. The order parameter critical exponent $\beta$ for
$\dot{\gamma}=1/32$ is similar to the value calculated for the Ising
model, but the values of  $z$, $\nu_{\parallel}$ and $\nu_{\bot}$
show that the anisotropy introduced by such a small external field
is relevant. At large $\dot{\gamma}$'s, all the exponents are
similar within a small range, suggesting that the critical behavior
of the model is practically independent of the magnitude of the
field in this regime. Furthermore, Table 3 also shows that
 the critical exponents of the
sheared model do not belong to the universality class of the Ising
or of the DLG model, even if this model shows a similar  phase
behavior.\\

\noindent Finally, the critical exponents summarized above were computed based
on the fact that only the longitudinal correlation length is
relevant for the dynamic critical behavior of the model,
independently of the initial configuration. In order to check this, 
a finite-size scaling of the dynamic evolution of $OP$ was performed
with the aid of equation (\ref{f-v}).
 This equation must contain
both the critical dynamic behavior of equations (\ref{preop}) and
(\ref{gsc1}) at early times of evolution, and also the finite-size
critical behavior at long times. As a consequence, the finite-size
exponents must be $\omega_{FDC}=\beta/\nu_{\parallel}$ and
$\omega_{GSC}=\beta/\nu_{\perp}$ for  both initial
conditions respectively. By measuring the saturated values of $OP$,
$OP_{sat}$, and computing
 the exponents $\omega_{FDC}$ and $\omega_{GSC}$
 the time series of $OP$ were collapsed for the system initiated from ordered and disordered
configurations as the main plots of figures \ref{satgd05} and
\ref{satgd10} show. Therefore, it is concluded
that only the longitudinal correlation length
$\xi_{\parallel}$ takes part in the critical evolution of the Ising
model when an external shear field is applied. Furthermore,
the finite-size exponents
 $\omega_{FDC}$ and $\omega_{GSC}$ were not consistent with the rates
  $\beta/\nu_{\parallel}$ and $\beta/\nu_{\perp}$ for the cases
   corresponding to $\dot{\gamma}=1/2$, while they are in good agreement
    for a shear field of magnitude $\dot{\gamma}=10$.
This discrepancy between the predicted and measured critical
exponents for the case $\dot{\gamma}=1/2$, together with the fact
that the values of the critical exponents estimated for smaller
$\dot{\gamma}$ are not similar with those corresponding to larger
values of $\dot{\gamma}$ (see Table \ref{tablacc}), may be explained
by conjecturing that, at such small values of the shear rate, the
system is less perturbed by the external driving. This means that
the growth of transverse critical correlations is less affected by
the shear field, and may become relevant in the short time regime.
If this is so, our scaling assumptions will be not longer valid, and
both $\xi$'s need to be considered in order to propose scaling forms
for the dynamic critical behavior of the model in this regime. In order to study this, new simulations of the model with $\dot{\gamma}=1/32$ were performed,
but they demanded a lot of computational time, specially when the system is started
from the GSC configurations because they needed larger lattice sizes and evolution time intervals in order to obtain good power laws and saturation regimes ($L\geq 10000$, evolution time intervals of the order of $10^6$ MCS or larger), so we did not obtained reliable results. As a consequence, this interesting subject will be left for further research in the future. In contrast, at larger $\dot{\gamma}'s$, the good agreement between $\beta/\nu_{\parallel}$ and $\beta/\nu_{\perp}$, calculated from the exponents in Table \ref{tablacc}, and the estimated values of $\omega_{FDC}$ and $\omega_{GSC}$ respectively, suggest that the critical behavior is different from both the cases with smaller $\dot{\gamma}'s$, and  also from the Ising and DLG models at large external fields \cite{alsa, repalsa}.

\section{Acknowledgements}

GPS wants to thank CONICET and the ANPCyT.

\end{document}